# Spatially reconfigurable topological textures in freestanding antiferromagnetic nanomembranes

Hariom Jani,[1,2]✉ Jack Harrison,[1] Sonu Hooda,[3] Saurav Prakash,[2] Proloy Nandi,[2] Junxiong Hu,[2] Zhiyang Zeng,[1] Jheng-Cyuan Lin,[1] Ganesh ji Omar,[2] Jörg Raabe,[4] Simone Finizio,[4]✉ Aaron Voon-Yew Thean,[3,5] A Ariando,[2,5]✉ Paolo G Radaelli[1]✉

[1] Clarendon Laboratory, Department of Physics, University of Oxford, Oxford, UK

[2] Department of Physics, National University of Singapore, Singapore

[3] Department of Electrical and Computer Engineering, National University of Singapore, Singapore

[4] Swiss Light Source, Paul Scherrer Institut, Villigen PSI, Switzerland

[5] Integrative Sciences and Engineering Programme, National University of Singapore, Singapore

✉ Correspondence to: hariom.jani@physics.ox.ac.uk; simone.finizio@psi.ch; ariando@nus.edu.sg; paolo.radaelli@physics.ox.ac.uk;

## Abstract

Antiferromagnets hosting real-space topological spin textures are promising platforms to model fundamental ultrafast phenomena and explore spintronics. However, to date, they have only been fabricated epitaxially on specific symmetry-matched crystalline substrates, to preserve their intrinsic magneto-crystalline order. This curtails their integration with dissimilar supports, markedly restricting the scope of fundamental and applied investigations. Here, we circumvent this limitation by designing detachable crystalline antiferromagnetic nanomembranes of α-$Fe_2O_3$, that can be transferred onto other desirable supports after growth. We develop transmission-based antiferromagnetic vector-mapping to show that these nanomembranes harbour rich topological phenomenology at room temperature. Moreover, we exploit their extreme flexibility to demonstrate three-dimensional reconfiguration of antiferromagnetic properties, driven locally via flexure-induced strains. This allows us to spatially design antiferromagnetic states outside their typical thermal stability window. Integration of such freestanding antiferromagnetic layers with flat or curved nanostructures could enable spin texture designs tailored by magnetoelastic-/geometric-effects in the quasi-static and dynamical regimes, opening new explorations into curvilinear antiferromagnetism and unconventional computing.

## Main Text

Topological textures in antiferromagnetic (AFM) materials are whirling structures with spins oppositely aligned between ferromagnetic (FM) sublattices. Beyond topological protection, this spin configuration affords unique benefits not enjoyed by their FM counterparts, including robustness against external perturbations, size scalability, and ultrafast dynamics.[1-5] In fact, some topological AFM textures are predicted to exhibit spintronic analogues of relativistic physics, where their speed limit is set by the magnon group velocity.[2,6] This immense potential has resulted in a surge of interest in topological AFM states.[1,3,7-9]

Central to nucleating and controlling topological textures are various magneto-crystalline interactions, namely anisotropy, exchange, and relativistic spin-orbit torques, which are intrinsic to specific antiferromagnets. AFM systems thus far reported to host topological order,[10-13] such as α-$Fe_2O_3$, CuMnAs, and $MnSc_2S_4$, were either bulk crystals[13] or were grown epitaxially on symmetry-matched crystalline substrates through advanced fabrication.[10-12] This markedly restricts their utility and flexibility in comparison to typical FM-based, topological texture hosting metallic heterostructures, which are polycrystalline and can be grown simply by sputter-deposition.[1,8] Therefore, further exploration and exploitation of topological AFM textures necessitates the development of crystalline AFM layers that can be readily integrated with dissimilar supports, for example, silicon or even non-crystalline flexible substrates.

To this effect, we drew inspiration from recent developments in crystalline quantum materials membranes, which are freestanding crystals of macroscopic lateral dimensions with a thickness of ~1-100 nm.[14-16] These membranes are a relatively new form of crystalline matter occupying an intermediate position between bulk and 2D materials, whilst having properties that are distinct from both. Generally, crystalline membranes have bulk-like magnetic/electronic properties, but, similar to 2D materials, are quite flexible, and can therefore withstand extreme deformations without undergoing fracture.[17,18] They can also be transferred post-growth to any desirable host, including silicon[15,19] or flexible supports,[20,21] enhancing the ability to stack and twist complex heterostructures.[22]

Here, we demonstrate the design and fabrication of high-quality AFM nanomembranes that preserve the all-important magneto-crystalline interactions post delamination. To image the local AFM order, we developed a scanning transmission X-ray microscopy (STXM)-based Néel vector reconstruction technique and demonstrated that our detached membranes host a multi-chiral family of topological AFM textures, analogous to the Kibble-Zurek-like phenomenology previously observed in attached epitaxial films.[10] Moreover, we provide a striking demonstration that the flexibility of our membranes can be exploited to drive a local reconfiguration of the AFM properties and topological textures across membrane 'folds'. We present a set of mechanical models, which demonstrate that our observations are consistent with the magneto-structural effects expected from flexure-based strains. Our results pave the way for the development of AFM spintronics platforms exploiting membrane tunability via geometry and strain.

## Membrane design and fabrication

We fabricated high-quality freestanding membranes of (001)-oriented, Rh-doped α-Fe$_2$O$_3$ (referred to hereafter as α-Fe$_2$O$_3$) using the *selective water-etching* technique[14,21,23] (see Methods) on epitaxial heterostructures grown by pulsed laser deposition.[10,24] Achieving high-quality α-Fe$_2$O$_3$ was found to depend critically on the choice of substrate and intermediate (buffer) layers to reduce inter-layer lattice mismatch. Due to the trigonal symmetry of α-Fe$_2$O$_3$ (space group $R\bar{3}c$), we chose single-crystalline (001)-oriented α-Al$_2$O$_3$ and (111)-oriented SrTiO$_3$ (STO) substrates as the growth templates, and (111)-oriented Sr$_3$Al$_2$O$_6$ (SAO) as the water-soluble sacrificial layer.[14,21,23] For all samples, water-etching of SAO resulted in freestanding oxide membranes, which were shifted to the desired support via either *direct* or *indirect transfer*, see Figure 1a. The former involves direct scooping of the floated membrane onto the support, whereas the latter requires spin-coating of a temporary organic support to hold the delaminated membrane before its final transfer. We have used both approaches for different experiments throughout this work.

We found that direct growth of α-Fe$_2$O$_3$|SAO on (001)-oriented α-Al$_2$O$_3$ substrates (sample *type-A*) results in oriented polycrystalline samples due to the large lattice mismatch between various layers in the stack, see Supplementary-S1. Film quality improves notably when α-Fe$_2$O$_3$|SAO are grown on (111)-oriented STO substrates (sample *type-B*) due to the significantly lower mismatch between SAO and STO (in this orientation in bulk, $a_{\text{SAO}}/4 \sim 5.60$ Å, $a_{\text{STO}} \sim 5.51$ Å), although the resulting α-Fe$_2$O$_3$ itself remains quite defective. To improve the sample quality further, we added an intermediate *buffer* consisting of an ultra-thin STO (111) layer followed by a thicker LaAlO$_3$ (LAO) (111) layer between SAO and α-Fe$_2$O$_3$ (sample *type-C*), see Methods and Supplementary-S1. Here, LAO acts as a good buffer as it has a slightly smaller lattice constant ($a_{\text{LAO}} \sim 5.35$ Å), reducing the mismatch with α-Fe$_2$O$_3$ ($a_{\text{Fe}_2\text{O}_3} \sim 5.03$ Å), whilst being structurally close to both STO and SAO.[23] Moreover, the ultra-thin STO increases the overall crystallinity[23] and is found to be critical in aiding the delamination of the overlayers in our buffered heterostructures. The addition of the buffer layers results in freestanding AFM membranes with much larger crack-free areas compared with the unbuffered counterparts, see Figure 1b and Supplementary-S1. Lastly, the sample quality and yield of these membranes are significantly superior to hematite layers prepared via chemical exfoliation.[25]

The quality and orientation of our buffered α-Fe$_2$O$_3$ crystal membranes were ascertained by X-ray diffraction (XRD) and selected-area electron diffraction (SAED), see Figure 1c-f, and Supplementary-S1. A unique feature of buffered α-Fe$_2$O$_3$ membranes is the formation of a moiré pattern evident in the reciprocal space as satellite peaks in SAED. This is expected to be a 'mismatch' moiré pattern,[26,27] which results from electron beam interference through the slightly mismatched lattices of α-Fe$_2$O$_3$ and the buffer layers. This picture is validated by our diffraction simulation, which closely reproduces the experimental pattern (Figure 1f). The resulting periodic perturbation at the α-Fe$_2$O$_3$-buffer interface does not appear to affect the magnetic properties of α-Fe$_2$O$_3$, as the length scales we study in magnetometry and X-ray microscopy, and the membrane thickness are significantly larger than those of the mismatch pattern.

## Magnetic transition in membranes

The reliable generation of topological textures in α-Fe$_2$O$_3$ requires the presence of a spin reorientation (Morin) phase transition, which mimics the Kibble-Zurek phenomenology.[10] At the Morin transition temperature $T_{\text{M}}$, the anisotropy undergoes a sign reversal,[10,24] causing spins to flip from out-of-plane (OOP) to in-plane (IP) configurations. The presence of a distinct Morin transition in the proximity of room temperature was confirmed both by SQUID magnetometry[10,24] and by dichroic X-ray spectroscopy, see Supplementary-S1. This is in sharp contrast to chemically exfoliated hematene membranes, which do not display any Morin transition.[25] Crucially, the transition in our detached membranes is qualitatively similar to those reported in attached epitaxial films,[10,24] despite the former being more defective than the latter, with transitions in buffered α-Fe$_2$O$_3$ being particularly sharp, see Supplementary-S1. We conclude that our water-etched membranes are good freestanding platforms to seek out real-space topological AFM order.

## Nanoscale mapping of the AFM order

To image the local AFM textures, we performed scanning transmission X-ray microscopy (STXM) in X-ray magnetic linear dichroism (XMLD) modality (see Methods) – an element-specific spectro-microscopy technique, with a large depth of focus, that enables unambiguous identification of the AFM contrast. In XMLD-STXM, a beam of Fe L$_3$-edge X-rays is focussed onto the AFM membranes at normal incidence, whilst changes in absorption are monitored in transmission by a point detector (Figure 2a). In this geometry, the X-ray polarisation (Linear Horizontal - LH) is in the basal plane of the α-Fe$_2$O$_3$ membranes, and IP and OOP AFM orientations are clearly distinguished as they contribute different XMLD contrast signals.[10] Moreover, by varying the sample azimuth using an *in situ* rotation stage, we changed the relative orientation of the X-ray polarization and IP Néel order systematically, enabling the nanoscale reconstruction of the AFM order.[10,28,29] Akin to our previous work with XMLD photoemission microscopy (PEEM),[10,28] the XMLD contrast can resolve the IP AFM directions but cannot distinguish the absolute sign of the AFM order. Nevertheless, we can clearly identify topological textures in our membranes from these reconstructions.

## Evolution across the Morin transition

XMLD-STXM reveals that our α-Fe$_2$O$_3$ membranes host magnetic textures remarkably similar to those seen in attached films (Figure 2b and Supplementary-S1).[10] For $T < T_M$, we observe large OOP AFM domains (purple) separated by antiphase domain walls (ADWs) with IP AFM order (yellow/orange). As the system is warmed, the ADWs widen and small IP islands nucleate and progressively increase in size. At $T \sim T_M$, the ADW length-scale diverges, as anisotropy approaches zero, resulting in a complex distribution of AFM domains hosting nearly equal fractions of IP and OOP regions. At $T > T_M$, IP regions enlarge and become dominant, whilst OOP regions shrink dramatically. Nonetheless, we still observe several OOP regions across the sample.

To determine the topological character of the AFM textures in our membranes we constructed Néel vector-maps for $T > T_M$. We used red-green-blue colours to denote IP domains with spin directions at 120° from each other, as expected based on the underlying trigonal symmetry.[28] Based on previous work in attached films,[10] we expect topological textures to be associated with small OOP 'cores'. Although such small topological cores are usually difficult to detect with LH-polarized X-rays, we observed some larger OOP 'bubbles', not all of which are associated with a whirling texture, and are therefore likely to be topologically trivial and produced by pinning of the OOP phase at local defects. More importantly, we were able to observe many topological AFM textures, including AFM merons and antimerons, see Figure 2c. Individual AFM (anti)merons can be characterised by an *AFM winding number* $\pm 1$ (depending on whether the texture whirls along or opposite to the azimuth angle), and an *AFM topological charge* $\pm 1/2$ (depending on the product of the winding number and the core polarisation).[4,9,10,28] Moreover, (anti)merons can combine locally to form pairs, which may have a net AFM topological charge of $\pm 1$ (bimerons) or $0$ (topologically trivial pairs), depending on the relative core polarisation of the (anti)merons. It should be noted that bimerons and topologically trivial pairs cannot be distinguished by XMLD techniques.[10] The observation of a multi-chiral topological AFM family unequivocally confirms that our membranes harbour the Kibble-Zurek phenomenology originally discovered in attached films,[10] despite the larger concentration of structural defects. Noteworthily, topological states are observed in the absence of any high spin-orbit heavy-metal overlayer, indicating that the creation and stabilization of topological order are intrinsic to α-Fe$_2$O$_3$, and do not rely on interfacial interactions present in typical FM skyrmionic systems.[1,5,8,9]

A clear difference between membranes and attached films is that AFM textures in the former are more strongly pinned than in the latter, so that texture patterns are reproduced almost identically even after performing multiple thermal cycles across $T_M$ (Supplementary-S2). We hypothesise that texture pinning results from highly localised alteration of the magnetic properties due to an increased density of point and extended defects in the membranes. This reasoning is supported by previous studies performed extensively in other topological systems.[30-33] We also performed *in situ* imaging with magnetic fields and found the AFM state to remain largely unperturbed (Supplementary-S2), indicating that our topological textures are much more robust compared to counterparts observed in synthetic AFMs.[5]

## Flexure-driven three-dimensional texture reconfiguration

One of the most remarkable properties of freestanding crystalline membranes is their extreme mechanical flexibility,[17,18,21] which could provide a powerful tuning 'handle' for systems with magnetic properties that are strain/structure sensitive.[34-36] We find that our buffered α-Fe$_2$O$_3$ membranes are not brittle, as one expects of ceramic-like oxides, but are very flexible and can develop 'folds'. An example is illustrated in Figure 3, where the fold has a maximum curvature of ~ 3×10$^5$ m$^{-1}$. In extreme scenarios, we even observe complete 180° 'folded-over' membranes, see Supplementary-S3. Large-area buffered membranes (type-C) are particularly remarkable, as they can hold complex strain distributions without undergoing fracture.

To study magneto-structural effects, we imaged naturally flexed regions across the membrane folds that emerged serendipitously upon direct transfer. Their shape was confirmed through confocal microscopy, which maps the height profile of the membrane (Figure 3b and Methods). Moreover, the slopes of the flexed region appear darker in STXM images (Figure 3c) because the signal scales inversely with the effective sample thickness $t_{\text{eff}} \sim t/\cos\theta$ ($t$ - actual thickness, $\theta$ - deviation angle from horizontal).

Flexure effects are immediately apparent from the images collected through the Morin transition. We define $T_M^{\text{FF}}$ as the Morin transition in the far-field (flat) region of the membrane away from the fold. All the data in Figure 3 were collected for a type-C membrane with the α-Fe$_2$O$_3$-layer facing upward, and the buffer layer lying underneath (for characterization details see Methods). Above $T_M^{\text{FF}}$ (Figure 3d), both the far-field regions and the peak of the fold exhibit an IP AFM matrix hosting several OOP cores, as expected. However, narrow bands near the base of the fold host a much large fraction of OOP AFM regions. Upon cooling to $T \sim T_M^{\text{FF}}$ (Figure 3e) the far-field regions exhibit mixed IP and OOP contrast, the base of the fold has a robust OOP matrix with clear AFM ADWs, whilst the top of the fold surprisingly remains in the IP state. Finally, at temperatures well below $T_M^{\text{FF}}$, all regions are in the OOP state interspersed with ADWs (Figure 3f). This remarkable evolution, consistent with the magnetic anisotropy changing rapidly through the fold, is in stark contrast with the behaviour in flat membranes, corresponding to Morin temperatures significantly *raised* near the base and *lowered* at the peak.

To further explore these effects, we imaged a buffered membrane in the reversed configuration, with the buffer layer facing upward, and the α-Fe$_2$O$_3$-layer is underneath, see Supplementary-S4. Here, the fold has the opposite effect with respect to the trend in Figure 3: the OOP state is stabilised on top of the fold (consistent with a *raising* of $T_M$), while the topologically-rich IP state is stabilised at the base (*lowering* of $T_M$). Finally, to confirm the presence of topological textures on folds, we also performed vector-mapping and found a clear instance of a meron-antimeron pair near the base. This suggests that structural reconfiguration could enable new form of spatial control over AFM topology.

The overall thermal evolution across different folded membranes can be summarised as follows: flexure alters the magnetic anisotropy in opposite directions near the peak and the base, such that the respective local $T_M$ is either suppressed or elevated relative to $T_M^{\text{FF}}$, but with the overall sign of the effect being reversed depending on whether α-Fe$_2$O$_3$ is the top or the bottom layer. Hence, we observe in Figure 3 that $T_M^{\text{peak}} < T_M^{\text{FF}} < T_M^{\text{base}}$, whereas in Supplementary-S4 we find $T_M^{\text{base}} < T_M^{\text{FF}} < T_M^{\text{peak}}$. These results indicate that magneto-structural effects can be used to spatially design AFM states outside their typical thermal stability window.

## Strain and Anisotropy Modelling

Magnetic anisotropy in α-Fe$_2$O$_3$ results from a delicate competition between dipolar and on-site interactions that are sensitive to structural variations.[24,34,37] In particular, epitaxial studies[34] revealed that uniform compressive and tensile strains applied via substrate clamping raise and lower $T_M$ in α-Fe$_2$O$_3$, respectively. Consequently, we seek to explain our experimental observations through flexure-induced strain.

To develop this simple insight, we estimated the strain distribution across our fold numerically through a finite-element mechanical model of a buffered α-Fe$_2$O$_3$ membrane (see Methods), whose flexed region closely reproduces the profile determined by confocal microscopy. We find that flexure results in sizable uniaxial in-plane tensile and compressive strains, $\varepsilon_{xx}$, distributed along the membrane thickness (z-direction), such that the *neutral* (unstrained) line[17] is located close to the middle of the buffered membrane. Due to the presence of the buffer, which itself accommodates some strain, the net strain $\langle\varepsilon_{xx}\rangle_{z,F}$ averaged across the α-Fe$_2$O$_3$ layer is actually non-zero. Moreover, $\langle\varepsilon_{xx}\rangle_{z,F}$ strength varies gradually across the length of the fold, changing sign near the point where the curvature is zero, see Figure 4a. Finally, this model predicts that reversing the buffered membrane should also reverse the sign of the strain distribution (Figure 4b).

To provide a semi-quantitative estimate of the flexure-induced variation of the magnetic anisotropy across the fold, we calculated the local $T_M$ (Figure 4c) by combining the thickness-averaged strain profile from our mechanical model with the strain dependence of $T_M$ determined in literature.[34] A caveat of this analysis is that the strains in Ref [34] were substrate-induced and biaxial, whereas the flexure-induced strains here are primarily uniaxial. In buffered membranes with the α-Fe$_2$O$_3$-side facing up, we find that the net compressive strain at the base of the fold ($\langle\varepsilon_{xx}\rangle_{z,F}^{\text{base}} < 0$) and the net tensile strain at the peak ($\langle\varepsilon_{xx}\rangle_{z,F}^{\text{peak}} > 0$) lead to an increase and decrease, respectively, of the local $T_M$ by ~10%. This is consistent with our experimental observations in Figure 3. Furthermore, the sign of this effect is reversed for the flipped buffered membrane, while its magnitude remains approximately the same, consistent with our results in Supplementary-S4.

As a final test of our magneto-structural model, we investigate flexed regions in unbuffered samples (type-B). Absence of the buffer layer should cause the neutral strain line in a folded membrane to lie approximately in the middle of the α-Fe$_2$O$_3$ layer, so that $\langle\varepsilon_{xx}\rangle_{z,F} \sim 0$. This would markedly suppress the flexure-driven anisotropy changes, in stark contrast to what is expected in buffered membranes. This is indeed confirmed by our experimental results illustrated in Supplementary-S5. We conclude that our model effectively explains the AFM state reconfiguration observed across the folded membranes.

## Discussion

Using a powerful transmission-based AFM vector-mapping technique, we demonstrated that high-quality α-Fe$_2$O$_3$ membranes host topological AFM states, which are reconfigurable via flexure-induced strain in three-dimensional folded structures.

At a fundamental level, our results establish that strain modulation is a novel and powerful tool to design and manipulate topological AFM textures, adding a completely new vista of reconfigurable magnetic topology to the burgeoning research landscape built on exploiting membranes of quantum materials to generate exotic states.[17,18,21] Our results also pave the way for the exploration of static and dynamical AFM evolution[38-41] triggered by *in situ* electric, magnetic, optical, or structural perturbations.[1,9] For example, we envisage electrically triggering topological reconfiguration and dynamics via localised piezoelectric control. Moreover, by integrating extremely flexible AFM membranes/ribbons onto carefully designed three-dimensional nanostructures, it may become possible to induce novel symmetry-breaking exchange or anisotropy interactions, e.g. through curvilinear geometric[35,36] and magnetoelastic[42,43] effects, thereby enabling the design of spatially-varying magnetic states,[44] or the realization of hitherto undiscovered chiral textures.[45,46]

On the applied front, the development of substrate-free AFM membranes that preserve magneto-crystalline properties and topological order addresses a major roadblock inhibiting the integration of crystalline AFM materials into established spintronics platforms. Specifically, complex and dense topological AFM fabrics are expected to possess fast non-linear dynamics,[2,47] which could open explorations into AFM-based silicon-compatible ultra-fast reservoir computing[7,48] or dense AFM memory-in-logic arrays in three dimensions.[49,50]

# Methods

**Membrane growth and fabrication:** Throughout this work, we have studied Rh-doped α-Fe$_2$O$_3$ (α-Fe$_{1.97}$Rh$_{0.03}$O$_3$) membranes. Rh-doping was used to elevate the Morin transition temperature beyond what is typically achievable in undoped counterparts, as discussed in the literature.[10,24] Membrane layers were grown either on (111)-oriented SrTiO$_3$ or (001)-oriented α-Al$_2$O$_3$ substrates from CrysTec, using a pulsed laser deposition setup fitted with a KrF excimer laser. Firstly, the growth of the water-soluble Sr$_3$Al$_2$O$_6$ layer was performed at 950 °C, in a pure oxygen atmosphere of 1 mTorr, and a laser repetition rate of 2 Hz. Deposition of the buffer SrTiO$_3$ layer (thickness ~ unit cells) was then performed at 850 °C, 10 mTorr, and 2 Hz. This was followed by the LaAlO$_3$ layer (~10 nm) grown at the same temperature and repetition rate, with an oxygen pressure of 1 mTorr. Subsequently, the α-Fe$_2$O$_3$ layer (~30 nm) was deposited at 700 °C, 2 mTorr, and 3 Hz. Finally, the samples were gradually cooled in a high-oxygen-pressure environment to minimise oxygen vacancies formed during the growth.

To delaminate the membranes,[14] samples were placed in high-purity deionised water at room temperature to dissolve the SAO layer. In the case of direct transfer (see Figure 1a), the membranes gradually floated to the top of the water surface, following which they were scooped out using the desired support (Si, Si$_3$N$_4$, etc.). The presence of non-uniform forces in the scooping process can result in the serendipitous formation of folded regions. These flexed geometrical structures are held in place due to van der Waals interactions with the underlying Si$_3$N$_4$ support. Alternatively, in the case of indirect transfer, a temporary support consisting of Poly(methyl methacrylate) (PMMA) was spin-coated on the top of the sample, which was then held by a flexible tape. The entire stack was then placed in deionised water. After delamination, the membrane was carefully moved to the final support (Si, Si$_3$N$_4$, etc.) using a transfer stage that held onto the tape. Lastly, the PMMA layer and the tape were removed from the top surface of the membrane through a room-temperature acetone wash. The indirect process enables targeted and controlled transfer of membranes with a much higher yield compared to its direct counterpart.

**Materials characterization**: The structural quality and crystallinity of the samples were determined by XRD involving $2\theta - \omega$ scans, rocking curves ($\omega$ scan), $\phi$ scans, and pole figure measurements, see Figure 1 and Supplementary-S1. Measurements were performed for both as-deposited films on crystalline substrates and membranes transferred onto Si substrates. The structural phase of α-Fe$_2$O$_3$ was further confirmed through Raman spectroscopy, performed using a Jobin Yvon Horiba LabRAM Evolution Spectrometer in reflection geometry (514.5 nm laser). Transmission electron microscope based SAED experiments were carried out using a JEM-ARM200F JEOL equipped with a cold field emission gun, operated at 200 kV. To ensure electron transparency, the AFM membranes were mounted on commercial 30-nm Si$_3$N$_4$ holders fabricated on top of Si substrates from Agar Scientific Ltd. Magnetic characterization was performed using a Quantum Design MPMS SQUID system on field-cooled samples under a 5000 Oe field during warming and cooling measurement scans. Detached membranes were supported on Kapton tape for magnetometry. Although there is generally good correspondence between magnetometry and STXM imaging, in some cases, the temperature dependence may have a minor difference, most likely due to small strain variation introduced during the corresponding sample preparation steps. The shape and height profile of membrane folds was studied through confocal microscopy using the Sensofar S Neox metrology tool.

**STXM imaging:** Fe L$_3$-edge resonant STXM imaging was performed at the PolLux X07DA endstation of the Swiss Light Source.[51] Images were obtained by recording the transmission of normally incident X-rays, polarised linearly (XMLD-STXM) or circularly (XMCD-STXM), and focused using a Fresnel zone plate (FZP). The outermost zone width of the FZP was selected, in tandem with the size of the monochromator exit slits of the beamline, resulting in a spatial resolution of about 40-50 nm. As the focusing efficiency of a diffractive optical element is not unitary, an order-sorting aperture (OSA), combined with a centre stop fabricated on the FZP, was employed to guarantee that only the focused X-rays illuminate the sample. The FZP and OSA are indicated in Figure 2a. An image is then obtained by scanning the sample with a piezoelectric scanner and recording the transmitted X-ray intensity for each point in the image. Typically, the field of view of our images had a square or rectangular shape.

For all X-ray transmission experiments, the AFM membranes were mounted on commercial 100-nm $Si_3N_4$ holders fabricated on top of Si substrates from Silson Ltd. The membrane temperature was controlled using a thin Au/Ti heater coil, lithographically fabricated directly onto the $Si_3N_4$/Si holders. Passing a current through the heater coil leads to resistive dissipation, thereby heating the sample. The resistance of the heater is calibrated and can be simultaneously measured to monitor the sample temperature. The Au/Ti heaters can be seen in Supplementary-S3,S4. For *in situ* field studies, magnetic fields were applied in the plane of the samples (i.e., in the crystal basal planes) using a rotatable permanent magnet setup, which could apply fields up to ~ 120 mT, see Supplementary-S2.

XMLD-STXM imaging was performed by collecting a pair of images, using linear horizontal X-ray polarization at two photon energies near the Fe L-edge (around ~ 710 eV), $E_1$ and $E_2$, chosen to provide maximal AFM contrast in our samples. The XMLD energy contrast was then calculated from $\Delta = (I_{E_1,LH} - I_{E_2,LH})/(I_{E_1,LH} + I_{E_2,LH})$. Based on crystal symmetry analysis,[10] it can be shown that the LH XMLD intensity varies as $I = I_A + I_B \cos^2 \psi$, where $\psi$ is the relative angle between the linearly polarised X-ray electric field and the magnetization, allowing us to map out the local AFM order. This immediately reveals that XMLD imaging is unable to distinguish IP AFM orientations separated by 180°. Likewise, it is also not possible to resolve the direction of OOP AFM orientation.[10,28] As the predominant AFM textures vary with temperature, the intensity range from the energy contrast also changes. Hence, we have used the same colour scheme (purple to yellow/orange for OOP to IP) with different energy contrast limits to aid the visualization across the transition.[10] Next, XMCD-STXM images presented in Supplementary-S1 were acquired by taking a set of data with both right (RCP) and left circularly polarised (LCP) X-rays at a fixed energy $E_i$. The XMCD contrast was calculated as $\delta = (I_{E_i,RCP} - I_{E_i,LCP})/(I_{E_i,RCP} + I_{E_i,LCP})$. Owing to the absence of strong ferromagnetic textures in the sample, XMCD-STXM images showed negligible contrast. The data reduction was performed using a custom-built MatLab tool (available on the public repository https://gitlab.psi.ch/microspectro-public/).

Spatially averaged X-ray absorption spectra were obtained at the Fe L-edges from the transmitted signal by scanning a straight line that straddled across regions both on and off the membrane, see Supplementary-S1. The signal measured outside the membrane was used as the reference signal for the normalization of the spectra acquired on the α-$Fe_2O_3$.

The depth of focus of the FZP utilized for the experiments reported in this work is ~ 1 μm, meaning that the imaging of the folded membranes had to be performed in several steps, bringing different parts of the fold into focus. Finally, composite images of AFM textures across the folds were produced by 'stitching' together multiple images.

To study the effects of flexure-induced strain in buffered membranes, it was crucial to determine whether α-$Fe_2O_3$ was on the top or the bottom of the stack, as non-uniform forces introduced during the scooping process can flip the membrane. To make this determination, we performed depth sensitive local elemental mapping, which could be accomplished by either of the two following techniques: (i) energy dispersive spectroscopy in transmission electron microscopy, or (ii) STXM imaging performed in total electron yield detection geometry, see Supplementary-S3. The latter was performed using a channeltron detector biased to a voltage of 2.4 kV. Here, only the secondary electrons emitted by the first few monolayers at the surface of the membrane are detected, in contrast to a typical transmission measurement where the entire thickness of the sample is probed. This allows us to determine the orientation of the membrane, depending on whether chemical contrast is detected at the Fe $L_3$-edge or the La $M_5$-edge, respectively, as shown in Supplementary-S3.

**Néel vector-maps:** The IP Néel vector-maps in Figure 2c and Supplementary-S4, which provide orientational information of the IP AFM order, were constructed by combining the energy-contrast XMLD-STXM images obtained at six azimuthal sample rotation angles about the crystallographic c-axis in the range -45° to +45° where the limits were set by *in situ* rotation stage, see Figure 2a. Theoretical and experimental details supporting this approach to Néel vector reconstruction can be found in our previous studies.[10,28,29] For each pixel in the field of view, we fit the angular dependence of the XMLD (see the previous section) to extract the average spin direction. Owing to the trigonal symmetry and the weak basal-plane anisotropy of α-$Fe_2O_3$, we mapped the spin directions using a red–green–blue (R-G-B) colour scale, which indicates the IP directions of the AFM order. For easy identification, the local IP Néel vector direction is indicated using a thin white bar. Regions in these maps where the AFM axis was identified to be OOP were coloured white.[10]

**Modelling and Simulation:** A time-dependent finite element analysis was carried out to study the strain profile of folded membranes quantitatively. The simulated composite stack consisted of both the buffered membrane (30 nm α-$Fe_2O_3$/10 nm LAO) and the 100 nm thick $Si_3N_4$ support. Young's moduli and Poisson's ratios of the materials used in the simulation are based on the reported values in the literature.[52-54] The simulation starts with the composite stack lying flat at rest, to which an impulsive upward external force is applied in the middle of the membrane causing a fold to emerge. Right after the removal of the external force, the two ends of the folded membrane are rigidly fixed to the $Si_3N_4$ support, with their separation distance corresponding to the experimental results from confocal microscopy (Figure 3b). Finally, both the membrane and the support deform and relax to their respective equilibrium states within ~1 μs simulation time. The resulting final state was found to closely reproduce the equilibrium state of a folded membrane held on the $Si_3N_4$ support. The only tuning parameter in the simulation is the length of the suspended composite membrane at rest, which is determined by matching the height of the folded membrane in the simulation to that in the experiment.

The SAED patterns were simulated by implementing the description of interfering diffraction patterns from two overlapping lattices, as developed in Refs.[26,27] The spatial moiré pattern is obtained as the sum of the two lattice functions, and the diffraction pattern is the Fourier transform of this combined function. The in-plane lattice parameters were chosen to correspond to experimental values of about 5.10 Å for α-$Fe_2O_3$ and 5.51 Å for LAO. For simplicity, we did not include the ultra-thin STO layer in this simulation as it is much thinner than the LAO in our buffered membranes.

# Figure 1

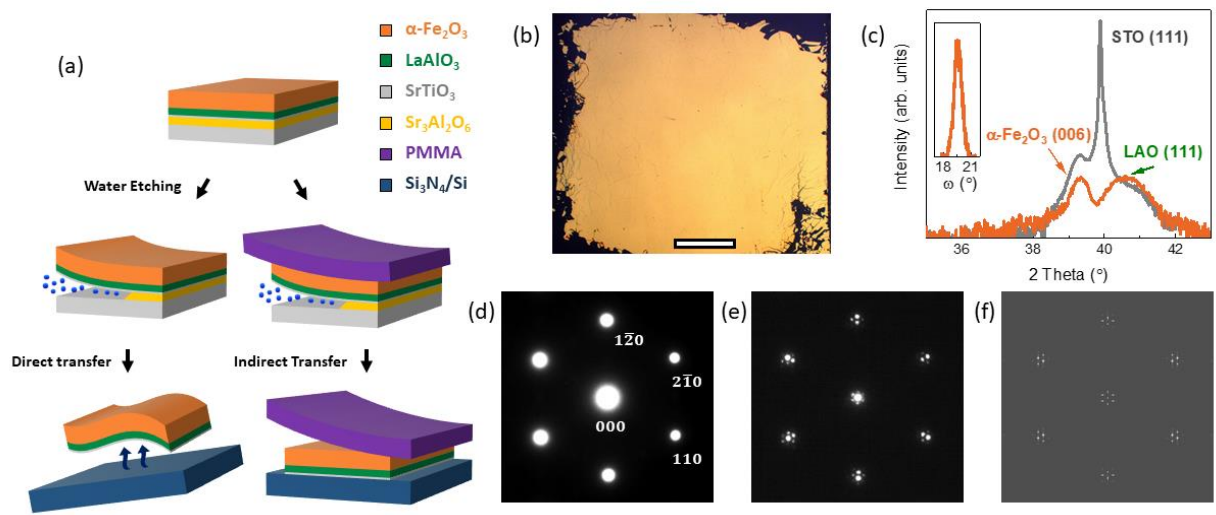

**Figure 1: Membrane design and characterization.** (**a**) Freestanding membranes are prepared by selective water-etching of SAO (yellow), followed by direct/indirect transfer of the membranes (orange) onto desired silicon (Si) or silicon nitride ($Si_3N_4$) supports (blue). Indirect transfer requires an intermediate support such as Poly(methyl methacrylate) (PMMA) (purple) to hold the membranes after water etching. For type-C membranes, a buffer layer made of LAO (green) and ultra-thin STO (grey) was also grown (see Methods). Layer thickness is not to scale. (**b**) Large-area optical image of a buffered α-$Fe_2O_3$ membrane transferred onto Si, with a spatial scale bar of 1 mm. (**c**) X-ray diffraction ($2\theta - \omega$ scans) of as-grown α-$Fe_2O_3$|LAO|STO|SAO film on an STO substrate (grey curve) and detached α-$Fe_2O_3$|LAO|STO membrane on $SiO_2$/Si substrate (orange curve). The out-of-plane (006) Bragg peak of α-$Fe_2O_3$ lies in the proximity of (111) LAO buffer and (111) STO substrate peaks. The STO layer in the buffer is too thin to contribute a sizable signal in the detached sample. The inset displays the rocking curve ($\omega$ scan) of the detached membrane, exhibiting a full-width half-maximum ~1.1°. (**d,e**) SAED patterns of free-standing (d) unbuffered (type-B) and (e) buffered (type-C) α-$Fe_2O_3$ membranes performed with the electron beam incident along the crystallographic *c*-axis. (**f**) Simulated SAED pattern of the type-C membrane (see Methods) corresponding to the pattern in (e). The simulation confirms that the satellite peaks in (e) emerge due to a lattice mismatch moiré pattern,[26,27] resulting from the electron beam interference across α-$Fe_2O_3$ and LAO lattices in the buffered membrane.

**Figure 2**

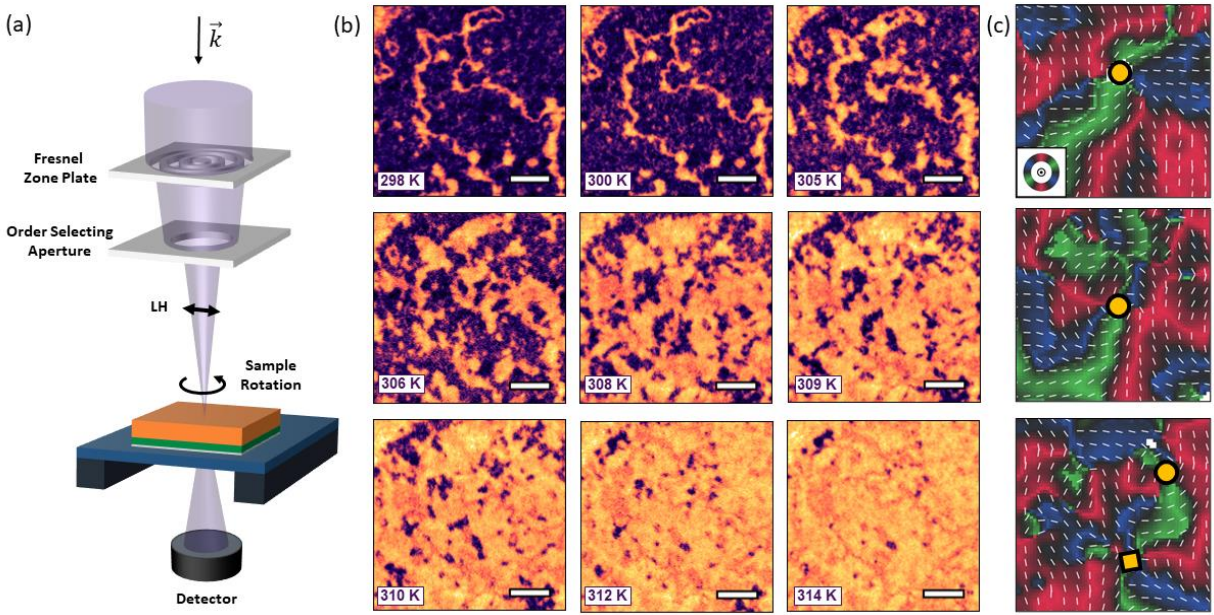

**Figure 2: Morin transition and generation of topological AFM textures. (a)** The geometry of the STXM measurement, performed using linearly polarised X-rays ($\vec{k}$) that are normally incident onto the sample. **(b)** Temperature evolution of the AFM STXM contrast obtained at the Fe $L_3$-edge while warming the buffered membrane (type-C) across $T_M$, in the temperature range 298 K – 314 K. The OOP and IP contrasts are indicated in purple and yellow/orange, respectively. The spatial scale bar is 2 µm long. All images were recorded at the same position. The energy contrast scale was varied slightly across the transition to aid visualisation.[10] **(c)** Vector-mapped STXM images performed at 314 K ($T > T_M$), produced by rotating the sample azimuth as shown in panel (a). R-G-B colours (key inset in c) and thin white bars represent the IP AFM orientations. White regions represent OOP orientations, whereas black regions highlight IP AFM directions deviating substantially from the R-G-B directions. Yellow circles and squares indicate AFM merons and antimerons, respectively. The width of the images is ~1.5 µm.

# Figure 3

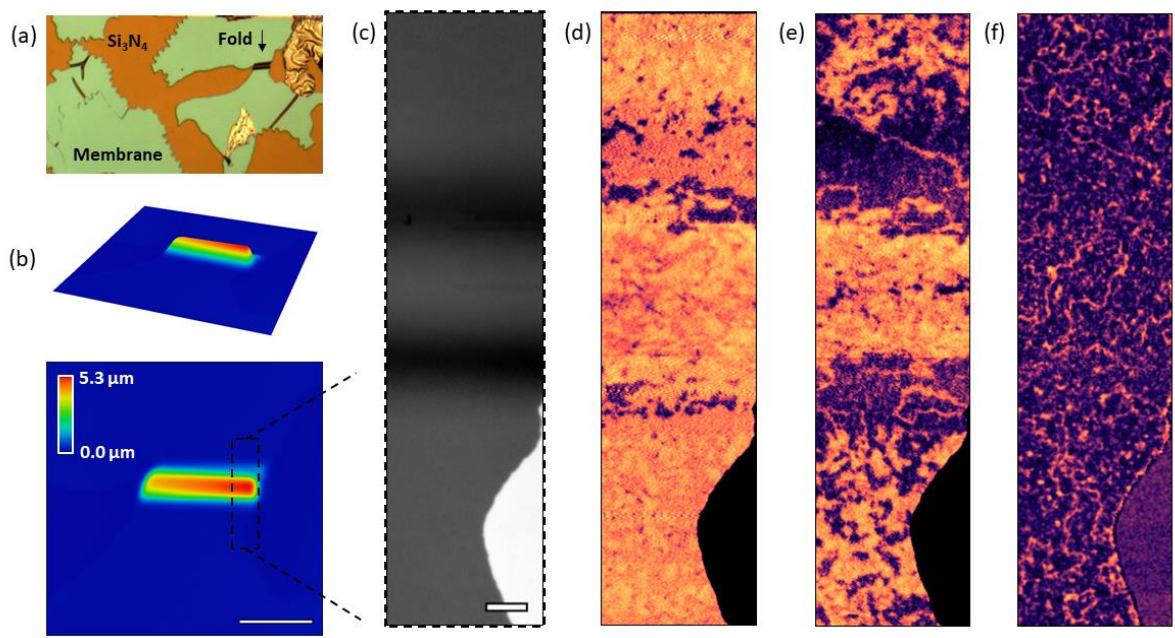

**Figure 3: Flexure-driven spatial reconfiguration of AFM textures.** (a) Optical microscopy image indicating the folded region being studied with a black arrow. (b) Three-dimensional height profile maps of a flexed region in a buffered membrane (type-C), shown from the side and top, obtained using confocal microscopy (see Methods). The colour legend indicates height profile. The spatial scale bar is 20 µm long. (c) Fe $L_3$-edge X-ray transmission contrast obtained at the right side of the folded membrane as indicated with a dashed box in (b). Our characterization confirmed that the buffered membrane was oriented with the α-$Fe_2O_3$-side facing up and buffer-side facing down (see Methods). The spatial scale bar is 2 µm long. (d,e,f) AFM STXM contrast obtained across different temperatures: (d) $T > T_M^{FF}$, (e) $T \sim T_M^{FF}$, and (f) $T < T_M^{FF}$. The OOP and IP contrasts are indicated in purple and yellow/orange, respectively, as in Figure 2. AFM textures across the folds were mapped out by 'stitching' together multiple images, whilst optimising the FZP focal point for each corresponding region.

**Figure 4**

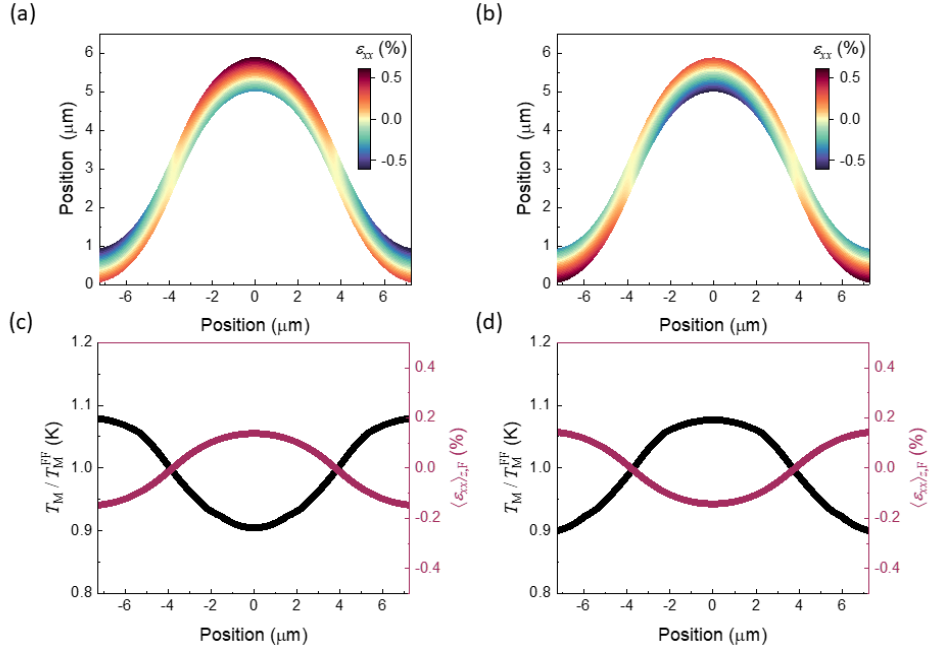

**Figure 4: Flexure strain and anisotropy model. (a,b)** Non-uniform strain distribution, $\varepsilon_{xx}$, in the α-Fe$_2$O$_3$ layer as a function of thickness and length, across folds in buffered AFM membranes (type-C) with the (a) α-Fe$_2$O$_3$-side and (b) buffer-side facing up configurations. Membrane thickness has been exaggerated to aid the visualization of the non-uniform strain distribution as a function of thickness. The neutral line ($\varepsilon_{xx} \sim 0$), indicated in yellow, is located at very different positions in (a) and (b) relative to the middle of the α-Fe$_2$O$_3$ layer, resulting in significant variation in the average and maximum strains built up in the AFM layer. **(c,d)** Evolution of the thickness averaged strain in the α-Fe$_2$O$_3$ layer, $\langle \varepsilon_{xx} \rangle_{z,F}$, and the corresponding local $T_\mathrm{M} / T_\mathrm{M}^\mathrm{FF}$ for the two configurations given in (a,b), respectively. Strain driven modulation of the local $T_\mathrm{M}$ was obtained from the model developed in the literature.[34] Here, $T_\mathrm{M} / T_\mathrm{M}^\mathrm{FF}$ larger and smaller than unity, refers to the elevation and suppression of the local Morin temperature and, therefore, the magnetic anisotropy, relative to the flat far-field regions.

## Acknowledgements

Work done at the University of Oxford was supported by Engineering and Physical Sciences Research Council (EPSRC) (EP/M020517/1) and the Oxford-ShanghaiTech collaboration project. Work done at the National University of Singapore was supported by the Agency for Science Technology & Research (A*STAR) under Advanced Manufacturing & Engineering Individual Research Grants (A1983c0034 and A2083c0054) and the Ministry of Education Academic Research Fund Tier 2 (MOE-T2EP50120-0015). H.J. acknowledges the support of Marie Skłodowska-Curie Postdoctoral Fellowship under the UK Research and Innovation Horizon Europe Guarantee Funding (EP/X024938/1). J.H. was supported by the EPSRC (DTP Grant No. 2285094). Z.Z. acknowledges the support of St. Peter's College Foundation Graduate Award. Part of this work was performed at the PolLux (X07DA) beamline of the Swiss Light Source, Paul Scherrer Institut, Villigen, Switzerland. PolLux endstation was financed by the German Bundesministerium für Bildung und Forschung (BMBF) under contracts 05K16WED and 05K19WE2. We acknowledge Singapore Synchrotron Light Source for time on the XDD beamline, which is supported by the National Research Foundation (Singapore). We thank T. Araki and B. Kaulich for supporting trial experiments at the Diamond Light Source (UK).

## Competing interests

The authors declare no competing interests.

## Author Contributions

H.J. performed materials design, growth and characterization. S.H., S.P. and J.X.H. developed membrane transfer protocols. H.J. and J.H. performed imaging experiments, with inputs from S.F. J.H. and H.J. performed data reduction and analysis. S.F. and J.R. developed STXM instrumentation. P.N. performed TEM experiments. J.H. performed diffraction simulations and vector reconstructions. Z.Z. developed the mechanical model. J.-C.L. and H.J. performed X-ray diffraction and magnetometry. A.A. supervised materials fabrication. H.J. and P.G.R. conceived and developed the project. All authors discussed the results and contributed to the manuscript.

## Supplementary Information

Section S1: Structural and magnetic characterization of membranes

Section S2: Robustness of AFM states

Section S3: Extreme membrane folds and their surface characterization

Section S4: Flexure effects on AFM textures in a flipped membrane

Section S5: Suppression of flexure effects in unbuffered membranes

# References

[1] Christian, H. B. *et al.* The 2020 Skyrmionics Roadmap. *Journal of Physics D: Applied Physics* **53**, 363001 (2020).
[2] Baltz, V. *et al.* Antiferromagnetic spintronics. *Reviews of Modern Physics* **90**, 015005 (2018).
[3] Büttner, F., Lemesh, I. & Beach, G. S. D. Theory of isolated magnetic skyrmions: From fundamentals to room temperature applications. *Scientific Reports* **8**, 4464 (2018).
[4] Barker, J. & Tretiakov, O. A. Static and Dynamical Properties of Antiferromagnetic Skyrmions in the Presence of Applied Current and Temperature. *Physical Review Letters* **116**, 147203 (2016).
[5] Legrand, W. *et al.* Room-temperature stabilization of antiferromagnetic skyrmions in synthetic antiferromagnets. *Nature Materials* **19**, 34-42 (2020).
[6] Shiino, T. *et al.* Antiferromagnetic Domain Wall Motion Driven by Spin-Orbit Torques. *Physical Review Letters* **117**, 087203 (2016).
[7] Grollier, J. *et al.* Neuromorphic spintronics. *Nature Electronics* (2020).
[8] Fert, A., Reyren, N. & Cros, V. Magnetic skyrmions: advances in physics and potential applications. *Nature Reviews Materials* **2**, 17031 (2017).
[9] Lim, Z. S., Jani, H., Venkatesan, T. & Ariando, A. Skyrmionics in correlated oxides. *MRS Bulletin* **46**, 1053-1062 (2021).
[10] Jani, H. *et al.* Antiferromagnetic half-skyrmions and bimerons at room temperature. *Nature* **590**, 74-79 (2021).
[11] Ross, A. *et al.* Structural sensitivity of the spin Hall magnetoresistance in antiferromagnetic thin films. *Physical Review B* **102**, 094415 (2020).
[12] Amin, O. *et al.* Antiferromagnetic half-skyrmions electrically generated and controlled at room temperature. *arXiv e-prints*, arXiv: 2207.00286 (2022).
[13] Gao, S. *et al.* Fractional antiferromagnetic skyrmion lattice induced by anisotropic couplings. *Nature* **586**, 37-41 (2020).
[14] Lu, D. *et al.* Synthesis of freestanding single-crystal perovskite films and heterostructures by etching of sacrificial water-soluble layers. *Nature Materials* **15**, 1255-1260 (2016).
[15] Han, L. *et al.* High-density switchable skyrmion-like polar nanodomains integrated on silicon. *Nature* **603**, 63-67 (2022).
[16] Kum, H. S. *et al.* Heterogeneous integration of single-crystalline complex-oxide membranes. *Nature* **578**, 75-81 (2020).
[17] Dong, G. *et al.* Super-elastic ferroelectric single-crystal membrane with continuous electric dipole rotation. *Science* **366**, 475-479 (2019).
[18] Xu, R. *et al.* Strain-induced room-temperature ferroelectricity in $SrTiO_3$ membranes. *Nature Communications* **11**, 3141 (2020).
[19] Bakaul, S. R. *et al.* Single crystal functional oxides on silicon. *Nature Communications* **7**, 10547 (2016).
[20] Li, C.-I. *et al.* van der Waal Epitaxy of Flexible and Transparent $VO_2$ Film on Muscovite. *Chemistry of Materials* **28**, 3914-3919 (2016).
[21] Hong, S. S. *et al.* Extreme tensile strain states in $La_{0.7}Ca_{0.3}MnO_3$ membranes. *Science* **368**, 71-76 (2020).
[22] Li, Y. *et al.* Stacking and Twisting of Freestanding Complex Oxide Thin Films. *Advanced Materials* **34**, 2203187 (2022).
[23] Chen, Z. *et al.* Freestanding crystalline $YBa_2Cu_3O_{7-x}$ heterostructure membranes. *Physical Review Materials* **3**, 060801 (2019).
[24] Jani, H. *et al.* Reversible hydrogen control of antiferromagnetic anisotropy in α-$Fe_2O_3$. *Nature Communications* **12**, 1668 (2021).
[25] Puthirath Balan, A. *et al.* Exfoliation of a non-van der Waals material from iron ore hematite. *Nature Nanotechnology* **13**, 602-609 (2018).
[26] Reidy, K. *et al.* Direct imaging and electronic structure modulation of moiré superlattices at the 2D/3D interface. *Nature Communications* **12**, 1290 (2021).
[27] Zeller, P. & Günther, S. What are the possible moiré patterns of graphene on hexagonally packed surfaces? Universal solution for hexagonal coincidence lattices, derived by a geometric construction. *New Journal of Physics* **16**, 083028 (2014).
[28] Chmiel, F. P. *et al.* Observation of magnetic vortex pairs at room temperature in a planar α-$Fe_2O_3$/Co heterostructure. *Nature Materials* **17**, 581-585 (2018).
[29] Waterfield Price, N. *et al.* Coherent Magnetoelastic Domains in Multiferroic $BiFeO_3$ Films. *Physical Review Letters* **117**, 177601 (2016).
[30] Holl, C. *et al.* Probing the pinning strength of magnetic vortex cores with sub-nanometer resolution. *Nature Communications* **11**, 2833 (2020).
[31] Gruber, R. *et al.* Skyrmion pinning energetics in thin film systems. *Nature Communications* **13**, 3144 (2022).

[32] Liang, X. *et al.* Dynamics of an antiferromagnetic skyrmion in a racetrack with a defect. *Physical Review B* **100**, 144439 (2019).
[33] Reichhardt, C., Reichhardt, C. J. O. & Milošević, M. V. Statics and dynamics of skyrmions interacting with disorder and nanostructures. *Reviews of Modern Physics* **94**, 035005 (2022).
[34] SeongHun, P. *et al.* Strain control of Morin temperature in epitaxial α-$Fe_2O_3$ (0001) film. *Europhysics Letters* **103**, 27007 (2013).
[35] Kravchuk, V. P. *et al.* Multiplet of Skyrmion States on a Curvilinear Defect: Reconfigurable Skyrmion Lattices. *Physical Review Letters* **120**, 067201 (2018).
[36] Makarov, D. *et al.* New Dimension in Magnetism and Superconductivity: 3D and Curvilinear Nanoarchitectures. *Advanced Materials* **34**, 2101758 (2022).
[37] Besser, P. J., Morrish, A. H. & Searle, C. W. Magnetocrystalline Anisotropy of Pure and Doped Hematite. *Physical Review* **153**, 632-640 (1967).
[38] Finizio, S., Mayr, S. & Raabe, J. Time-of-arrival detection for time-resolved scanning transmission X-ray microscopy imaging. *Journal of Synchrotron Radiation* **27**, 1320-1325 (2020).
[39] Donnelly, C. *et al.* Time-resolved imaging of three-dimensional nanoscale magnetization dynamics. *Nature Nanotechnology* **15**, 356-360 (2020).
[40] Büttner, F. *et al.* Observation of fluctuation-mediated picosecond nucleation of a topological phase. *Nature Materials* **20**, 30-37 (2021).
[41] Wintz, S. *et al.* Magnetic vortex cores as tunable spin-wave emitters. *Nature Nanotechnology* **11**, 948-953 (2016).
[42] Gomonay, H. & Loktev, V. M. Magnetostriction and magnetoelastic domains in antiferromagnets. *Journal of Physics: Condensed Matter* **14**, 3959-3971 (2002).
[43] Eliseev, E. A., Morozovska, A. N., Glinchuk, M. D. & Blinc, R. Spontaneous flexoelectric/flexomagnetic effect in nanoferroics. *Physical Review B* **79**, 165433 (2009).
[44] Donnelly, C. *et al.* Complex free-space magnetic field textures induced by three-dimensional magnetic nanostructures. *Nature Nanotechnology* **17**, 136-142 (2022).
[45] Harrison, J., Jani, H. & Radaelli, P. G. Route towards stable homochiral topological textures in A-type antiferromagnets. *Physical Review B* **105**, 224424 (2022).
[46] Pylypovskyi, O. V. *et al.* Curvilinear One-Dimensional Antiferromagnets. *Nano Letters* **20**, 8157-8162 (2020).
[47] Hals, K. M. D., Tserkovnyak, Y. & Brataas, A. Phenomenology of Current-Induced Dynamics in Antiferromagnets. *Physical Review Letters* **106**, 107206 (2011).
[48] Bourianoff, G., Pinna, D., Sitte, M. & Everschor-Sitte, K. Potential implementation of reservoir computing models based on magnetic skyrmions. *AIP Advances* **8**, 055602 (2018).
[49] Gu, K. *et al.* Three-dimensional racetrack memory devices designed from freestanding magnetic heterostructures. *Nature Nanotechnology* **17**, 1065 (2022).
[50] Luo, Z. *et al.* Current-driven magnetic domain-wall logic. *Nature* **579**, 214-218 (2020).
[51] Raabe, J. *et al.* PolLux: A new facility for soft x-ray spectromicroscopy at the Swiss Light Source. *Review of Scientific Instruments* **79**, 113704 (2008).
[52] Chicot, D. *et al.* Mechanical properties of magnetite ($Fe_3O_4$), hematite (α-$Fe_2O_3$) and goethite (α-FeO·OH) by instrumented indentation and molecular dynamics analysis. *Materials Chemistry and Physics* **129**, 862-870 (2011).
[53] Khan, A., Philip, J. & Hess, P. Young's modulus of silicon nitride used in scanning force microscope cantilevers. *Journal of Applied Physics* **95**, 1667-1672 (2004).
[54] Luo, X. & Wang, B. Structural and elastic properties of LaAlO3 from first-principles calculations. *Journal of Applied Physics* **104**, 073518 (2008).